\documentstyle[times,prl,aps,multicol,epsf]{revtex}
\begin{document}
\newcommand{\Dr}{\Delta_{\rm r}}
\newcommand{\eps}{\epsilon_0}
\newcommand{\eh}{\hat\epsilon}
\newcommand{\om}{\omega}
\newcommand{\omc}{\omega_{\rm c}}
\newcommand{\pmin}{p_{\rm min}}
\newcommand{\ple}{p_{\rm le}}
\newcommand{\pue}{p_{\rm ue}}
\newcommand{\pphi}{\tilde{p}_\phi}
\renewcommand{\epsilon}{\varepsilon}
\renewcommand{\phi}{\varphi}
\newcommand{\phit}{\phi_0}
\draft
\title{Stabilizing dynamical localization in driven tunneling}

\author{J\"urgen T. Stockburger and C. H. Mak}

\address{Department of Chemistry, University of
Southern California, Los Angeles, CA 90089-0482}

\date{Received 18 June 1998}

\maketitle

\begin{abstract}
Dynamical localization of a tunneling system by means of an
oscillating external field is examined for an arbitrary doublet of
tunneling states. We show that the condition of an exact crossing of
Floquet levels, required for localization in a symmetric system, can
be substantially relaxed for tunneling systems with broken
symmetry. This generalization equally applies to tunneling systems
coupled to a dissipative environment.
\end{abstract}


\begin{multicols}{2}
\setcounter{unbalance}{3}
\narrowtext A tunneling particle can be localized indefinitely by an
external periodic force.  This surprising prediction was made recently
by Grossmann {\it et al.} \cite{GrossPRL,GrossEPL}. With properly
chosen frequency and amplitude the external force can pin the
tunneling particle, and the amplitude of the tunneling oscillations
can be made arbitrarily small. Numerous methods have since been
applied to study this effect, focusing on special cases which lend
themselves to analytical solutions
\cite{LlorentePlata,Kayanuma,WangShao,WangZhao,YuriBavli93,%
YuriBavliMetiu96}, or on their generalizations to tunneling systems
with dissipation
\cite{DittrichOelschlaegelHaenggi,GrossmannDittrichJungHaenggi,%
Yuri,Milena95,Milena96,MilePeter,Nancy,Neu}. Previous works are based
on the central assumption that an exact crossing of Floquet levels is
necessary for dynamical localization to occur.  For Floquet levels to
cross, the tunneling system typically has to be symmetric, and the
driving amplitude must be selected from a set of discrete ``magic
numbers'', different for each driving frequency \cite{StrongBiasNote}.

In this Letter, we show that the tunneling particle can remain pinned
even if both of the above conditions are relaxed simultaneously.  This
implies that the class of systems in which the dynamical localization
can potentially be detected is much wider than previously thought.
These include tunneling systems which are not intrinsically symmetric,
such as tunneling of an atom between a sample and an STM
tip\cite{LouisSethna}, or tunneling in disordered systems like
structural glasses\cite{AHV,Phillips}. We demonstrate that these
findings also apply to dissipative tunneling systems, where the
localization effect manifests itself through a suppression of
tunneling oscillations and a drastic slowing down of the incoherent
dynamics.

To study driven tunneling systems, we can either use a double-well
potential model\cite{GrossPRL} or, more simply, a truncated 
two-state Hamiltonian \cite{GrossEPL}
\begin{equation}\label{H0}
H_0 = -{\hbar\Delta\over 2} \sigma_x - {\hbar\over 2}(\eps +
\eh\sin\om t)\sigma_z .
\end{equation}
This truncation is valid if all parameters are small
compared to the lowest oscillation frequency $\om_{\rm osc}$
associated with the double well, and for a barrier height larger than
$\hbar\om_{\rm osc}$. $\sigma_x$ and $\sigma_z$ are Pauli spin
matrices, $\hbar\Delta$ is the tunnel splitting, $\hbar\eps$ an
intrinsic energy bias, and $\hbar\eh\sin\om t$ its modulation by an
external periodic force.

Despite the simplicity of this Hamiltonian, its dynamics shows
surprising effects in parameter regimes for which the rotating-wave
approximation fails. A general analytical solution is not available,
hence a detailed understanding of its dynamics must rely on numerics.

The Hamiltonian (\ref{H0}) is non-conservative, but it exhibits discrete
time-translational invariance with period $T=2\pi/\om$. Consequently,
the one-period time translation operator $U_T = U(T,0)$ contains
all essential information about the dynamics. Previous studies 
\cite{GrossPRL,GrossEPL,LlorentePlata,Kayanuma,WangShao} concentrated on the
case where two Floquet states, the eigenstates of $U_T$ in the
two-state description, are degenerate. In that case, $U_T$ is the identity
operator (or its negative), and the time dependence of all observables
is strictly periodic with period $T$. Within the interval $T$, 
the population $p(t)$ of the initially occupied site exhibits 
small oscillations, with a minimum $\pmin$ such that $|1-\pmin| \ll 1$.

In this work, we will consider this inequality the definition of
localization and make no assumptions about the spectrum of $U_T$. In
general, $\pmin$ is to be defined as the largest lower bound (infimum)
of $p(t)$ for any positive $t$,
\begin{equation}
\pmin = \inf\limits_{t\geq 0} p(t) .
\end{equation}
Although $t$ is unbounded, $\pmin$ can be evaluated without
uncontrolled approximations or extrapolation. Using the group
properties of discrete time translations, the dynamics at arbitrary
time $t$ can be constructed from the time translation operator at
times $\tau\leq T$,
\begin{equation}
U_t = U_\tau U_T^n\;.
\label{deco}
\end{equation}
Here we have defined $U_t = U(t,0)$, with $t=nT+\tau$, $0\leq\tau<T$.
The lower bound $\pmin$ is thus given by
\begin{equation}
\pmin = \min\limits_{0\leq\tau\leq T} \inf\limits_{n\geq 0}
p_n(\tau) , \label{pmindble}
\end{equation}
where
\begin{equation}
p_n(\tau) = \left| \langle\psi_0 | U_\tau U_T^n | \psi_0\rangle
\right|^2
\end{equation}
generally defines an infinite set of functions on the interval
$0\leq\tau\leq T$.

To obtain the infimum with respect to $n$, it is useful to represent
the SU(2) matrix $U_T$ by a generator $g$ and an angle $\phit$,
$
U_T = \exp({\textstyle {i\over 2}} \phit g)$ .
%
The unit vector $\vec{n}$ in $g=\vec{n}\cdot\vec{\sigma}$ denotes the
rotation axis of the corresponding SO(3) rotation. Now $p_n$ reads
\begin{equation}
p_n(\tau) = \left| \langle\psi_0 | U_\tau
\exp({\textstyle {i\over 2}} n\phit g)
| \psi_0\rangle \right|^2 ,
\end{equation}
and one can define a quantity closely related to $p_n(\tau)$ with a
continuous index $\phi$,
\begin{equation}
\pphi(\tau) = \left| \langle\psi_0 | U_\tau \exp({\textstyle {i\over
 2}} \phi g) | \psi_0\rangle \right|^2 ,
\end{equation}
where $0\leq \phi \leq 4 \pi$. The set of functions $\{p_n(\tau)\}$
is a subset of $\{\pphi(\tau)\}$, i.e.,
\begin{equation}
\min\limits_{0\leq\phi\leq 4\pi} \pphi(\tau) \leq \inf\limits_{n\geq
0} p_n(\tau) .
\label{pn_leq_pphi}
\end{equation}
For generic parameters, $\phit/\pi$ is an irrational number. The set
$\{\exp({i\over 2} n\phit g)\,|\,n\geq 0\}$ is a dense subset of
$\{\exp({i\over 2} \phi g)\,|\,0\leq\phi\leq 4\pi\}$, and strict
equality holds in (\ref{pn_leq_pphi}). But in the degenerate case of
rational $\phit/\pi$, the situation is different. $p(t)$ is periodic,
and there are only a finite number of functions $p_n(\tau)$ that
are distinct. We shall consider the latter case irrelevant
in the long-time limit, because any minute drift or other
aberration in the driving frequency or amplitude will reduce it to the
generic case.  We will concentrate on the generic case in the 
following.

For the two-state system, $\pphi(\tau)$ is a continuous function of
$\phi$ and $\tau$, taking the form $\pphi(\tau) = a_\tau + b_\tau
\sin\phi + c_\tau \cos\phi$. The coefficients $a_\tau$, $b_\tau$, and
$c_\tau$ are algebraic functions of the matrix elements of $U_T$ and
$U_\tau$.  From the numerical solution of the Schr\"odinger equation
over a {\em single period} one can thus easily obtain the {\em entire
set} of functions $\pphi(\tau)$, and also the lower and upper
envelopes of the set of curves $p_n(\tau)$,
\begin{eqnarray}
\ple(\tau) &=& \inf\limits_{n\geq 0} p_n(\tau) 
 = \min\limits_{0\leq\phi\leq 2\pi} \pphi(\tau) \label{ple} ,\\
\pue(\tau) &=& \sup\limits_{n\geq 0} p_n(\tau)
 = \max\limits_{0\leq\phi\leq 2\pi} \pphi(\tau) \label{pue} .
\end{eqnarray}
According to (\ref{pmindble}), $\pmin$ is then given by the minimum of
the lower envelope,
\begin{equation}
\pmin = \min\limits_{0\leq\tau\leq T} \ple(\tau) \label{minle}.
\end{equation}

For a symmetric system driven at frequencies $\om\gg\Delta$, $\pmin$
has been shown to be almost unity if the ratio $\eh/\om$ belongs to a
set of discrete `magic numbers' for which $U_T$ is diagonal and its
eigenstates degenerate\cite{GrossPRL,GrossEPL}. In the
high-frequency limit, these numbers are the zeros $z_n$ of the Bessel
function $J_0(z)$. But with even a slight deviation from these numbers,
the degeneracy of the Floquet states will be lifted and $U_T$ will have {\em
delocalized} eigenstates, which manifest themselves as slow oscillations in
$p(t)$ with large amplitude and a reduced tunneling frequency 
$\Delta J_0(\eh/\om)$.

The central finding of this Letter is that the instability of the
localization effect can be suppressed by imposing a small static bias $\eps$
on the tunneling system. This
is demonstrated by the numerical data in Fig.\ 1a. For a symmetric system
(dashed curve), a slight detuning of the amplitude from its magic
value leads to to a slow coherent transition that depopulates the
initial state. Adding a small bias, however, prevents this depopulation even 
out to very long times (solid curve). Fig.\ 1b shows the corresponding set of
curves $p_n(\tau)$ for $0\leq n \leq 3$ as well as the lower and upper
envelopes $\ple(t)$ and $\pue(t)$. This rigorous result for $\ple$
shows unequivocally that a very robust localization persists
{\em indefinitely} for the detuned system. The periodic
function $p(t)$ for the corresponding symmetric system with perfectly
tuned amplitude $\eh/\om \approx 2.393$ is given for comparison (dashed
curve). Both cases show roughly equal $\pmin$.

How large does the bias need to be in order to effect dynamical localization, 
and is there a maximum allowable value for it?
The precise answer depends quite sensitively on the
driving parameters.  We can look for the answer using the seminumerical
procedure outlined above. The global minimum in Eq.\ (\ref{minle}) can
easily be obtained by linear search because $\ple(\tau)$ is smooth.
Fig.\ 2 presents quantitative results about the dependence of $\pmin$
on $\eps$ for different $\eh$ in a tunneling system driven at
$\om=5\Delta$. The width of the minimum at $\eps=0$ vanishes as
$\eh/\om$ approaches the localization point of the symmetric system,
i.e., the bias required to enable localization goes to zero. 
On the other hand, for very large bias $\eps \approx \om$, 
$\pmin$ drops drastically, and the dynamics is then characterized by large
oscillations\cite{YuriBavliMetiu96}.

A qualitative understanding of this generalized localization effect is
obtained by evaluating the dynamics at finite time analytically in the
limit of high frequency \cite{LlorentePlata,WangZhao}. For $\om \gg
\max(\eps,\Delta$), we find $|\phit |\ll 1$, and $g=N^{-1}(\Delta
J_0(\eh/\om)\sigma_x + \eps\sigma_z)$. From this we conclude that
localization is possible if either (i) $\phit = 0$ (Floquet level
crossing) or (ii) $|J_0(\eh/\om)| \ll |\eps/\Delta|$ (axis $\vec{n}$
aligned with the $z$ axis). The two cases are closely related. In either
case, driving modulates the transitions amplitudes giving rise to the
off-diagonal elements of $U_T$. In the first case, this is done in
such a way that destructive interference makes the
off-diagonal elements vanish precisely. 
In the second case, this cancellation is
imperfect, but the residual off-diagonal elements are small enough to
permit localized eigenstates of $U_T$.

In a condensed matter environment, tunneling is coupled to
fluctuations of the surrounding medium, and the delicate cancellations
underlying this localization effect are disturbed by dissipation and
dephasing. The standard approach to linear dissipation resulting from
such a coupling generalizes the two-state system to the driven
spin-boson Hamiltonian\cite{Leggett,Weiss},
\[
H = H_0 + \sum_\nu \omega_\nu (a^\dagger_\nu a_\nu
+ {\textstyle{1\over 2}} )
+ {q_0\over 2} \sum_\nu C_\nu (a_\nu + a^\dagger_\nu )\sigma_z.
\]
Here and in the following we set $\hbar=1$; $q_0$ is the distance
between tunneling sites. The effect of the harmonic
environment is fully characterized by the spectral density $ J(\om')
= \pi \sum_\nu C_\nu^2 \, \delta(\om'-\omega_\nu )$. Its specific
form depends on the density of states of the dissipative environment
and the details of their local interaction with the tunneling
system. The Ohmic form $J(\om') = \eta \om' \exp(-\om'/\omc)$ is one of
the most frequently investigated\cite{Leggett,Weiss}.
$q_0$ and $\eta$ appear in the dynamics only through the dimensionless
dissipation constant $\alpha = \eta q_0^2/2\pi$.  In the
case of large $\omc$ applicable to tunneling in solid-state systems,
$\Delta$ and $\omc$ enter only through the scaled tunneling frequency
$\Dr = \Delta (\Delta/\omc)^{\alpha/(1-\alpha)}$.

With the notable exception of the noninteracting-blip approximation
(NIBA)\cite{Leggett,Weiss} and certain limiting cases, the dynamics
governed by the driven spin-boson Hamiltonian appears to be
intractable without resorting, at least partially, to numerical methods.
The recently introduced chromostochastic quantum dynamics (CSQD)
algorithm\cite{CSQD} makes possible the
exact and efficient computation of all dynamical quantities of
interest in the spin-boson system. The CSQD method minimizes memory
effects inherent to quantum dissipation by direct sampling of the
colored noise produced by the environment. The statistical error
increases only moderately with $t$; in the examples given below it
never exceeds $7\times 10^{-3}$.

The dynamics at low temperature and weak damping closely resembles
that of the undamped system. Fig.\ 3 illustrates dynamical
localization in a weakly damped asymmetric tunneling system with
$\eh/\om = 2.3$ driven at $\om=5\Dr$. For a symmetric system, the same
parameters result in a coherent population transfer between the two
localized states. This transfer is strongly suppressed when a moderate
bias $\eps=-1$ is introduced.  The system now exhibits a gradual
population decay on a much longer timescale. Comparison with the free
population decay of the biased system (without driving) shows that the
bias $\eps$ alone is not sufficient to cause localization. In this
example, the tunneling particle is actually localized on the site
whose energy is {\em raised} by the bias. Fig.\ 4 shows the same
driven system with $\eps=-1$ (top) and $\eps=1$ (bottom) with little
difference between the two curves, reminiscent of the symmetry seen in
Fig.\ 2. The center plot shows a symmetric system driven at the same
frequency, but with the amplitude needed for a Floquet crossing,
$\eh/\om \approx 2.393$. Evidently, the two cases of localization
discussed above lead to nearly identical behavior also in the
dissipative case.

We have demonstrated that the remarkable effects resulting from
dynamical localization in tunneling systems occur for a wide range of
parameters that do {\em not} obey the restrictive conditions for an
exact Floquet level crossing. We believe that these findings should
stimulate further theoretical and experimental investigation. The
prospect of an experimental realization has improved appreciably. The
hard-to-satisfy requirements of an extremely precise control of the
driving amplitude and its perfect homogeneity throughout a sample can
be dropped. The range of possible applications is substantially widened
to include asymmetric and even disordered systems as
candidates.  Considering, e.g., the density and sound
velocity of vitreous silica and the strain coupling $\gamma\approx
2{\rm eV}$ of its intrinsic tunneling systems\cite{Graebner},
dynamical localization may be achieved by ultrasonic driving at low
temperature. At a frequency of 1~GHz, the required amplitude for
localization will be reached at an acoustic intensity of about
4~mW/cm$^2$. Another phenomenon recently predicted and proposed as a
testbed for theories of dissipative tunneling is the tunneling of a
single atom between an STM or AFM tip and a
sample\cite{LouisSethna}. The adsorbed atom forms dipoles of opposite
direction on the sample and tip surfaces and thus experiences a force
from an applied DC or AC tip voltage. For the Ni:Xe tunneling system
proposed in \cite{LouisSethna} and characterized in more detail in
\cite{WalkupEtAl}, the localization condition for a 10~GHz driving
frequency will be reached with an AC tip voltage of only about 2~mV.

This research has been supported by the National Science Foundation
under grants CHE-9257094 and CHE-9528121, by the Sloan Foundation
and by the Dreyfus Foundation. 
Computational resources have been provided by
the IBM Corporation under the SUR Program at USC. J.~S. enjoyed
stimulating discussions with M. Grifoni and U. Weiss.

\begin{figure}
\epsfxsize=1.0\columnwidth \centerline{\epsffile{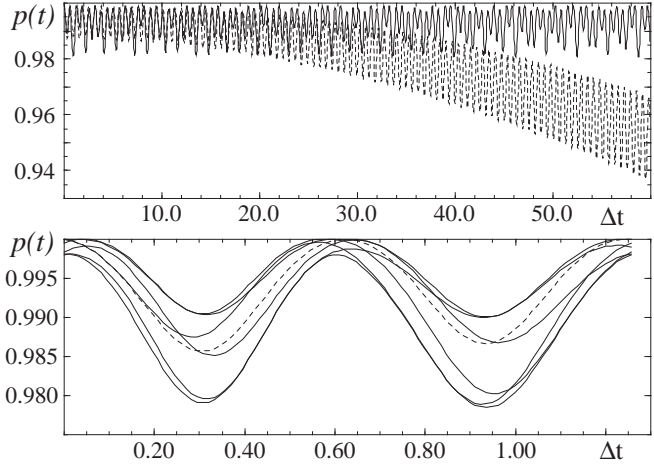}}
\caption[]{{\bf (a)} Time-dependent population $p(t)$ for symmetric
(dashed line) and asymmetric ($\eps=\Delta$, solid line) tunneling
systems with $\om = 5\Delta$, and $\eh/\om = 2.38$. {\bf (b)} {\em
Asymmetric} system (solid lines): Non-periodic trajectories $p_n(t)$
(thin) with periodic upper and lower envelope (thick) vs. $\Delta
t$. {\em Symmetric} system (dashed line): Periodic $p(t)$ of the
dynamically localized system at $\eh/\om \approx 2.393$.}
\end{figure}

\begin{figure}
\epsfxsize=0.9\columnwidth
\centerline{\epsffile{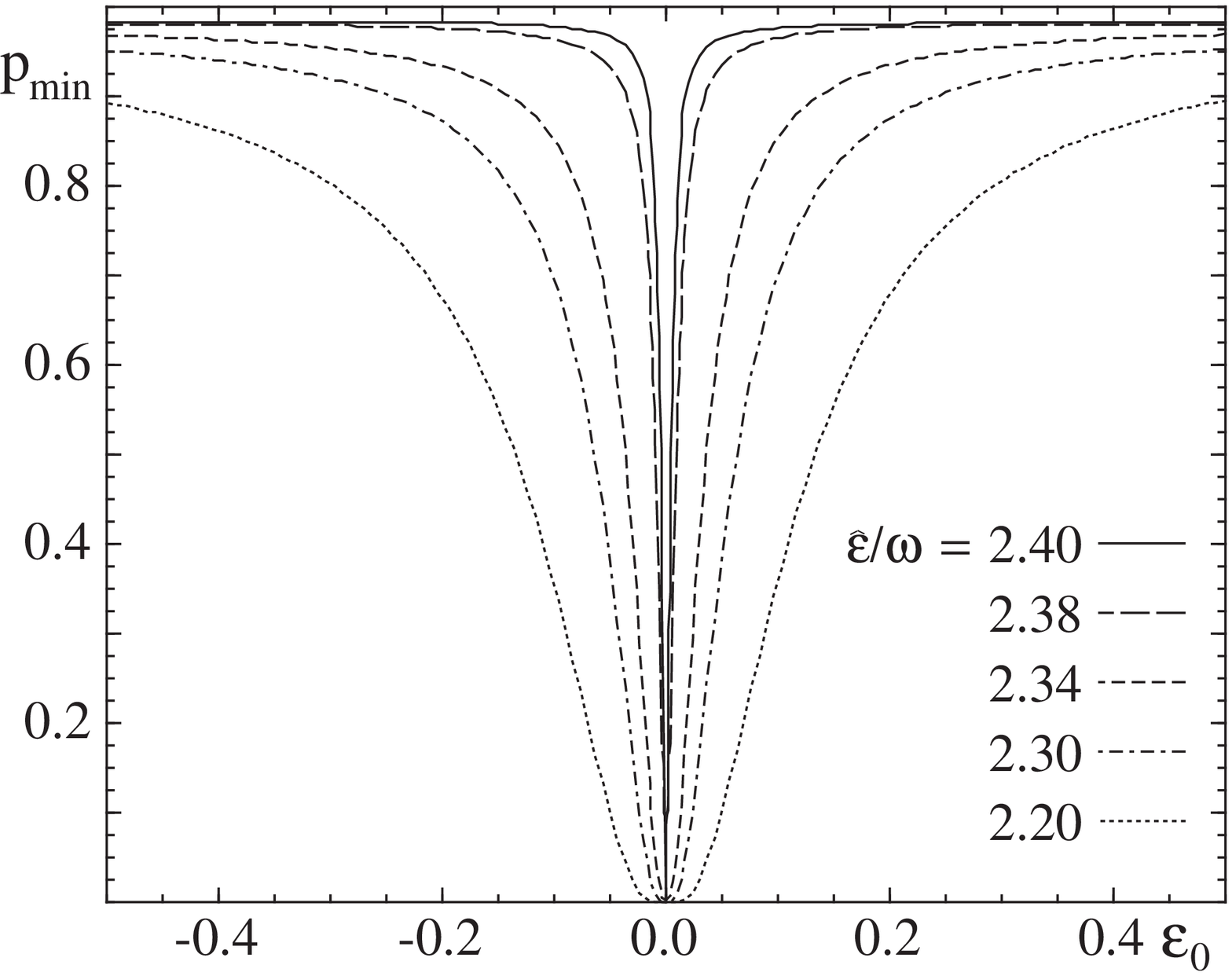}}
\caption[]{Lower population bound $\pmin$ vs. bias $\eps$ for various
ratios $\eh/\om$ at $\om=5$ (unit $\Delta=1$).}
\end{figure}

\begin{figure}
\epsfxsize=0.9\columnwidth
\centerline{\epsffile{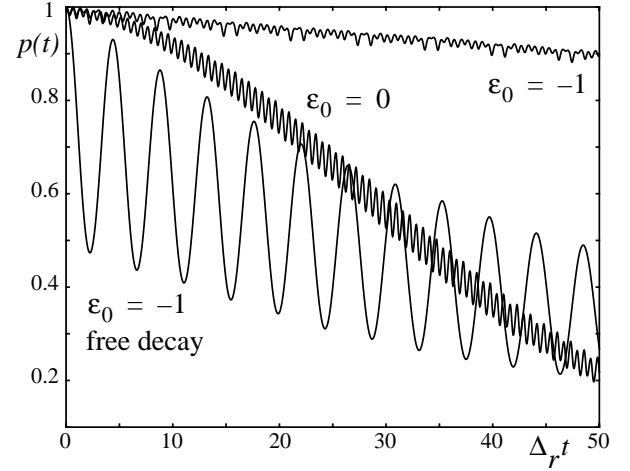}}
\caption[]{Population decay for weak dissipation ($\alpha=0.01$, zero
temperature) with driving at $\om=5\Dr$, $\eh/\om = 2.3$, and $\omc =
200\Dr$. Top curve -- biased system: localized, long population decay
time. Middle curve -- symmetric system: coherent population
transfer. Lower curve -- biased system, no driving: fast tunneling
oscillations and incoherent decay.}
\end{figure}

\begin{figure}
\epsfxsize=1.0\columnwidth
\centerline{\epsffile{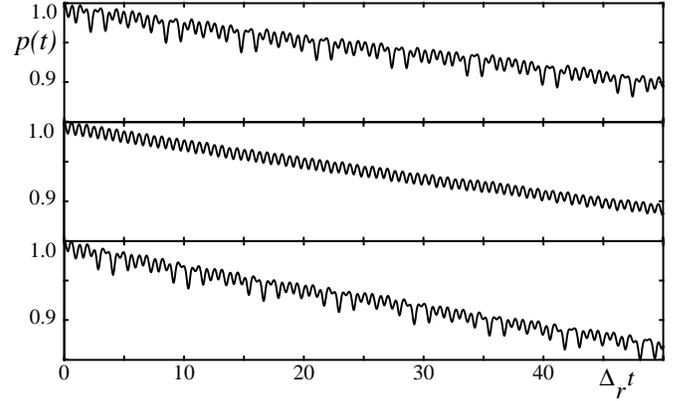}}
\caption[]{Slow population decay of a localized tunneling system
driven at $\om=5\Dr$ with weak dissipation ($\alpha = 0.01$, zero
temperature). Top: $\eps = -1$, $\eh/\om = 2.3$. Center: $\eps = 0$,
$\eh/\om \approx 2.393$. Bottom: $\eps = 1$, $\eh/\om = 2.3$.}
\end{figure}

\end{multicols}
\end{document}